\documentstyle[12pt,cite]{article}

\renewcommand{\today}{July 14, 1994}

\newcommand{\nc}{\newcommand}
\nc{\be}{\begin{equation}}
\nc{\ee}{\end{equation}}
\nc{\bea}{\begin{eqnarray}}
\nc{\eea}{\end{eqnarray}}
\nc{\beas}{\begin{eqnarray*}}
\nc{\eeas}{\end{eqnarray*}}
\nc{\noi}{\noindent}
\nc{\sD}{\not \! \! D}
\nc{\s}[1]{\not \! #1}
\nc{\non}{\nonumber}
\nc{\bb}{\bibitem}
\nc{\lf}{\left}
\nc{\r}{\right}
\nc{\mb}[1]{\makebox[#1]{}}
\nc{\pa}{\partial}
\nc{\sA}{\not \! \! A}
\nc{\newsec}[1]{\section{#1}\mb{0.5cm}}
\nc{\h}{\frac{1}{2}}
\nc{\ra}{\rightarrow}
\nc{\la}{\leftarrow}

\begin{document}
\thispagestyle{empty}
\begin{flushright}
ADP-94-7/T149 \\
hep-ph/9405273

\end{flushright}

\begin{center}
{\large{\bf Constraints on the momentum dependence of
 rho-omega mixing}}
 
 \vspace{1cm}
 {\large{\bf  [Phys. Lett. B336 (1994) 1-5]}} 

\vspace{2.2 cm}
H.B. O'Connell, B.C. Pearce, \\
 A.W. Thomas and A.G. Williams \\
\vspace{1.2 cm}
{\it
Department of Physics and Mathematical Physics \\
University of Adelaide, S.Aust 5005, Australia } \\
\vspace{1.2 cm}
\today
\vspace{1.2 cm}
\begin{abstract}
Within a broad class of models we show that the amplitude for $\rho^0-\omega$
mixing must vanish at the transition from timelike to spacelike four momentum.
Hence in such models the mixing is either zero everywhere
or is necessarily momentum-dependent.  This lends support to the conclusions
of other studies
of rho-omega mixing and calls into question standard
assumptions about the role of rho-omega
mixing in the theoretical understanding of charge-symmetry breaking in
nuclear systems.

\end{abstract}
\end{center}
\vspace{2.5cm}
\begin{flushleft}
E-mail: hoconnel, bpearce, athomas, awilliam@physics.adelaide.edu.au
\vfill
\end{flushleft}

\newpage


Charge symmetry violation (CSV) is a small but well established feature of the
strong nucleon-nucleon ($NN$) force \cite{a,b,c}. The class III force which
differentiates the $nn$ and $pp$ systems is best established through the
Okamoto-Nolen-Schiffer anomaly in the binding energies of mirror nuclei
\cite{d,e}. In the $np$ system the class IV CSV interaction mixes spin-singlet
and spin-triplet states. Despite presenting a difficult experimental challenge
this has been seen in high precision measurements at TRIUMF and IUCF
\cite{f,g}

Although there is still no universally accepted theoretical description of the
short and intermediate range $NN$ force, the one-boson-exchange model provides
a conceptually simple, yet quantitatively reliable framework \cite{h}. Within
that approach $\rho-\omega$ mixing is a major component of both class III and
class IV CSV forces \cite{a,c,j,k,l,l'}. For on-mass-shell vector mesons,
$\rho-\omega$ mixing is observed directly in the measurement of the pion
form-factor in the time-like region (that is, in the reaction
$e^+e^-\ra\pi^+\pi^-$ \cite{B}). The best value of the strong interaction
contribution to this amplitude at present is $\left<\rho^0|H_{\rm
str}|\omega\right>=-(5130\pm 600){\rm MeV}^2$ (on mass shell) from Hatsuda
{\it et al.} \cite{HHMK}.
 (A small, calculable, electromagnetic contribution of $\approx
610{\rm MeV}^2$ from $\rho\ra\gamma\ra\omega$ has been subtracted from the
data ($-4520\pm 600{\rm MeV}^2$) to leave the strong mixing amplitude.) Within
QCD this provides an important constraint on the mass differences of the $u$
and $d$ quarks
\cite{m}.

Of course, with respect to the CSV component of the $NN$ force a significant
extension is required. In particular, the exchanged vector meson has a
space-like momentum, far from the on-shell point. For roughly twenty years it
was customary to assume that the $\rho-\omega$ mixing amplitude was a constant
over this range of four-momentum. Only a few years ago Goldman, Henderson, and
Thomas (GHT) questioned this assumption \cite{GHT}. Within a simple model they
showed that the mixing amplitude had a node near $q^2=0$ so that neither the
sign nor magnitude in the spacelike region was determined by the on-shell
value. Since the initial work by GHT a qualitatively similar result has been
obtained using many theoretical approaches including mixing via an
$N\overline{N}$ loop using the $p-n$ mass difference \cite{PW}, several
$q\overline{q}$ calculations \cite{MTRC,p}, and an approach using QCD
sum-rules \cite{HHMK,H}. All of these calculations revealed a node at or near
$q^2=0$, with a consequent change in the sign of the mixing amplitude.  The
presence of this node in the corresponding coordinate space CSV potential has
been stressed in Refs.~\cite{HHMK,GHT,p,o}. Related studies of the
$\pi^0-\eta$ mixing have also been recently made including $N\overline{N}$
\cite{q} and $q\overline{q}$ \cite{r} loops, and chiral perturbation theory
\cite{s}. Significant momentum dependence was also observed in these studies.

It is important to note that the only calculation which found a node at
exactly $q^2=0$ was that of Piekarewicz and Williams \cite{PW}. In this work
alone was local current conservation guaranteed exactly. We have been led to
examine the general constraint on the mixing amplitude at $q^2=0$ by this
observation as well as by several inquiries from K. Yazaki \cite{n}. Our
findings can be summarised very easily. We argue that the
mixing amplitude vanishes at $q^2=0$ in any effective Lagrangian
model [e.g., ${\cal L}(\vec\rho,\omega,\vec\pi,\bar\psi,\psi,\cdots)$],
where there are no
explicit mass mixing terms [e.g., $M^2_{\rho\omega}\rho^0\omega$ or
$\sigma\rho^0\omega$ with $\sigma$ some scalar field]
in the bare Lagrangian and where the
vector mesons
have a local coupling to conserved currents which satisfy the usual vector
current commutation relations. The boson-exchange model of ref.~\cite{PW}
where, e.g., $J^\mu_\omega = g_\omega\bar N\gamma^\mu N$, is one particular
example. It follows
that the mixing tensor (analogous to the full
self-energy
function for a single vector boson such as the $\rho$ \cite{BLP})
\be
C^{\mu\nu}(q)=i\int d^4x\,e^{iq\cdot x}\left<0\right|T(J^\mu_{\rho}(x)
J^\nu_\omega(0))\left|0\right>.
\label{eq:tensor}
\ee
is transverse.
Here, the operator $J^\mu_\omega$ is the operator appearing in the
equation of motion for the field operator $\omega$, i.e., the Proca
equation given by $\partial_\nu F^{\mu\nu}-M^2_{\omega}\omega^\mu =
J^\mu_\omega$.  Note that when $J^\mu_\omega$ is a conserved current
then $\partial_\mu J^\mu_\omega=0$, which ensures that the Proca equation
leads to the same subsidiary condition as the free field case,
$\partial_\mu \omega^\mu=0$ (see, e.g., Lurie, pp.~186-190, \cite{spec}).
The operator $J^\mu_\rho$ is similarly defined.
We see then that $C^{\mu\nu}$ can be written in the form,
\be
C^{\mu\nu}(q)=\left(g^{\mu\nu}-{q^\mu q^\nu \over q^2}\right)C(q^2)\,.
\label{transverse}
\ee
{From} this it follows that the one-particle-irreducible self-energy or
polarisation, $\Pi^{\mu\nu}(q)$ (defined through eq.~(\ref{piC}) below),
 must also be transverse \cite{BLP}.
The essence of the argument below is that since
there are no massless, strongly interacting vector particles $\Pi^{\mu\nu}$
cannot be singular at $q^2=0$ and   therefore
$\Pi(q^2)$ (see eq.~(\ref{pi}) below)
must vanish at $q^2=0$, as suggested for the pure $\rho$ case
\cite{HFN}. As we have already noted this is something that was approximately
true in all models, but guaranteed only in Ref.~\cite{PW}.

Let us briefly recall the proof of the transversality of $C^{\mu\nu}(q)$.
As shown, for example, by Itzykson and Zuber (pp.~217-224) \cite{IZ},
provided we use covariant
time-ordering the divergence of $C^{\mu\nu}$ leads to a naive commutator of
the appropriate currents
\bea
\nonumber
q_\mu C^{\mu\nu}(q)&=&-\int d^4x\,e^{iq\cdot x}\pa_\mu \;[\theta(x^0)
\left<0\right|J^\mu_\rho(x)J^\nu_\omega(0)\left|0\right> \\
&\,\,&+\;\theta(-x^0)\left<0\right|J^\nu_\omega(0) J^\mu_\rho(x)
  \left|0\right>] \\
&=& -\int d^3x\,e^{i\vec{q}\cdot \vec{x}}
\left<0\right|[J^0_\rho(0,\vec{x}),J^\nu_\omega(0)]\left|0\right>_{\rm naive}.
\label{commutator}
\eea
That is, there is a cancellation between the seagull and Schwinger terms.
Thus, for any model in which the isovector- and isoscalar- vector currents
satisfy the same commutation relations as QCD we find
\be
q_\mu C^{\mu\nu}(q)=0.
\ee
Thus, by Lorentz invariance, the tensor must be of the form
given in eq.~(\ref{transverse}).

For simplicity we consider first the case of a single vector meson (e.g. a
$\rho$ or $\omega$) without channel coupling. For such a system one can
readily see that since $C^{\mu\nu}$ is
transverse the one-particle irreducible self-energy, $\Pi^{\mu\nu}$, defined
through \cite{BLP}
\be
\Pi^{\mu\alpha}D_{\alpha\nu}=C^{\mu\alpha} D^0_{\alpha\nu}
\label{piC}
\ee
(where $D$ and $D^0$ are defined below) is also transverse.
 Hence
\be
\Pi^{\mu\nu}(q)=\left(g^{\mu\nu}-{q^\mu q^\nu \over q^2}\right)\Pi(q^2)\,.
\label{pi}
\ee

We are now in a position to establish the behaviour of the
 scalar function, $\Pi(q^2)$. In
a general theory of massive vector bosons coupled to a conserved current, the
bare propagator has the form
\be
D^0_{\mu\nu}=\left(-g_{\mu\nu}+{q_\mu q_\nu \over M^2}\right){1 \over
{q^2-M^2}}\ee
whence
\be
(D^0)^{-1}_{\mu\nu}=(M^2-q^2)g_{\mu\nu}+q_\mu q_\nu.
\ee
The polarisation is incorporated in the standard way to give the dressed
propagator
\bea
\nonumber
D^{-1}_{\mu\nu}&=&(D^0)^{-1}_{\mu\nu}+\Pi_{\mu\nu} \\
&=&(M^2-q^2+\Pi(q^2))g_{\mu\nu}+\left(1-{\Pi(q^2)\over q^2}\right)q_\mu q_\nu.
\label{eq:inverse}
\eea
Thus the full propagator has the form
\be
D_{\mu\nu}(q)={-g_{\mu\nu}+\left(1-[\Pi(q^2)/q^2]\right)(q_\mu q_\nu/
M^2)\over q^2-M^2-\Pi(q^2)}.
\label{prop1}
\ee
Having established this form for the propagator, we wish to compare it with
the spectral representation of the propagator \cite{IZ,R,spec},
\be
D_{\mu\nu}(q)
=-i\int_{r_0}^\infty dr{\rho(r) \over {q^2-r}}\left(g_{\mu\nu}-{q_\mu q_\nu
\over r}\right).
\ee
Since no massless states exist in the strong-interaction sector we must have
$r_0>0$. Hence it is a straightforward exercise to show that we can write for
some function $F(q^2)$ \cite{R}
\be
D_{\mu\nu}(q)= F(q^2)g_{\mu\nu}+{1\over q^2}(F(0)-F(q^2))q_\mu q_\nu.
\label{prop2}
\ee
By comparing the coefficients of $g_{\mu\nu}$ in eqs.~(\ref{prop1}) and
(\ref{prop2}) we deduce
\be
F(q^2)={-1\over {q^2-M^2-\Pi(q^2)}}\,,
\ee
while from the coefficients of $q_\mu q_\nu$ we have
\bea
\nonumber
{\left(1-{\Pi(q^2)\over q^2}\right)\over {(q^2-M^2-\Pi(q^2))M^2}}&=&{1\over
q^2}(F(0)-F(q^2)) \\
&=& {1\over q^2}{q^2 +\Pi(0)-\Pi(q^2) \over (M^2+\Pi(0))(q^2-M^2-\Pi(q^2))},
\eea
from which we obtain
\be
\frac{\Pi(0)}{q^2}(q^2-M^2-\Pi(q^2))=0 \,\,\,\,\,\,\,\, ,\,\,\,\,\,\forall q^2
\ee
and thus
\be
\Pi(0)=0.
\ee
This embodies the principal result of the investigation, namely that
$\Pi(q^2)$ should vanish as $q^2\ra 0$ at least as fast as $q^2$.

While the preceding discussion dealt with the single channel case, for
$\rho-\omega$ mixing we are concerned with two coupled channels. Our
calculations therefore involve matrices. As we now demonstrate, this does not
change our conclusion.

The matrix analogue of eq.~(\ref{eq:inverse}) is
\be
D^{-1}_{\mu\nu}=\left(\begin{array}{cc}
M_\rho^2g_{\mu\nu}+(\Pi_{\rho\rho}(q^2)-q^2)T_{\mu\nu} &
\Pi_{\rho\omega}(q^2)T_{\mu\nu} \\
\Pi_{\rho\omega}(q^2)T_{\mu\nu} & M_\omega^2g_{\mu\nu}+
(\Pi_{\omega\omega}(q^2)-q^2)T_{\mu\nu}
\end{array} \right),
\ee
where we have defined $T_{\mu\nu}\equiv g_{\mu\nu}-(q_\mu q_\nu/q^2$)
for brevity.
By obtaining the inverse of this we have the two-channel propagator
\be
D_{\mu\nu}=\frac{1}{\alpha}\left(\begin{array}{cc}
s_\omega g_{\mu\nu}+a(\rho,\omega) q_\mu q_\nu &  \Pi_{\rho\omega}(q^2)
T_{\mu\nu} \\
 \Pi_{\rho\omega}(q^2)T_{\mu\nu} & s_\rho g_{\mu\nu}+a(\omega,\rho) q_\mu q_\nu
\end{array} \right),
\label{prop3}
\ee
where
\bea
s_\omega&\equiv&q^2-\Pi_{\omega\omega}(q^2)-M_\omega^2 \\
s_\rho &\equiv&q^2-\Pi_{\rho\rho}(q^2)-M_\rho^2 \\
a(\rho,\omega)&\equiv&\frac{1}{q^2 M_\rho^2}\{\Pi_{\rho\omega}^2(q^2)-
[q^2-\Pi_{\rho\rho}(q^2)]s_\omega\} \\
\alpha &\equiv& \Pi_{\rho\omega}^2(q^2)-s_\rho s_\omega .
\eea
In the uncoupled case [$\Pi_{\rho\omega}(q^2)=0$] eq.~(\ref{prop3}) clearly
reverts to the appropriate form of the one particle propagator,
eq.~(\ref{prop1}), as desired.

We can now make the comparison between eq.~(\ref{prop3}) and the Renard form
\cite{R} of the propagator, as given by eq.~(\ref{prop2}). The transversality
of the off-diagonal terms of the propagator, demands that
$\Pi_{\rho\omega}(0)=0$. A similar analysis leads one to conclude the same for
$\Pi_{\rho\rho}(q^2)$ and $\Pi_{\omega\omega}(q^2)$.
Note that the physical $\rho^0$ and $\omega$ masses which arise from locating
the poles in the diagonalised propagator matrix $D^{\mu\nu}$ no longer
correspond to exact isospin eigenstates.  To lowest order in CSV the
physical $\rho$-mass is given by $M^{\rm phys}_\rho
=[M^2_\rho+\Pi_{\rho\rho}((M^{\rm phys}_\rho)^2)]^{1/2}$, i.e., the pole
in $D^{\mu\nu}_{\rho\rho}$.
The physical $\omega$-mass is similarly defined.

In conclusion, it is important to review what has and has not been
established. There is no unique way to derive an effective field theory
including vector mesons from QCD. Our result that $\Pi_{\rho\omega}(0)$ (as
well as $\Pi_{\rho\rho}(0)$ and $\Pi_{\omega\omega}(0)$) should vanish
applies to those effective theories in which: (i) the vector mesons have local
couplings to conserved currents which satisfy the same commutation relations
as QCD [i.e., eq.~(\ref{commutator}) is zero]
and (ii) there is no explicit mass-mixing term in the bare
Lagrangian. This includes a broad range of commonly used, phenomenological
theories.
It does not include the model treatment of Ref.~\cite{MTRC} for example,
where the mesons are bi-local objects in a truncated
effective action.  However, it is interesting to note
that a node near $q^2=0$ was found in this model in any case.
The presence of an explicit mass-mixing term in the
bare Lagrangian will shift the
mixing amplitude by a constant (i.e., by $M_{\rho\omega}^2$).  We
believe that such a term will lead to difficulties in matching
the effective model
onto the known behaviour of QCD in the high-momentum limit, \cite{newOPTW}.

Finally the fact that $\Pi(q^2)$ is momentum-dependent or vanishes everywhere
in this class of models implies that the conventional {\it assumption}
of a non-zero, constant $\rho-\omega$ mixing amplitude remains
questionable.  This study then lends support to
those earlier calculations, which we briefly discussed, where it was
concluded that the mixing may play a minor role in the explanation of CSV
in nuclear physics. It remains an
interesting challenge to find possible alternate mechanisms to describe
charge-symmetry violation in the $NN$-interaction
\cite{la,lb}.

\vspace{1.5cm}
{\bf Acknowledgments}

One of us (AWT) would like to thank Prof. K. Yazaki for several stimulating
discussions.  This work was supported by the Australian Research Council.

\end{document}